\documentclass[referee]{raa}
\usepackage{graphicx,times}
\usepackage{natbib}
\usepackage{amssymb,amsmath}
\bibpunct{(}{)}{;}{a}{}{,}

\usepackage[a4paper=true,dvipdfm=true,pagebackref=true]{hyperref}
\hypersetup{pdftitle = The title of my PDF, pdfauthor = My name, pdfsubject= The subject, pdfkeywords = keyword1 keyword2 keyword3}
\hypersetup{colorlinks = true, linkcolor = green, anchorcolor = red, citecolor = blue, filecolor = red, pagecolor = red, urlcolor = red}

\begin{document}

   \title{The first photometric analysis and period investigation of the K-type W UMa type binary system V0842 Cep
}

 \volnopage{ {\bf 20XX} Vol.\ {\bf X} No. {\bf XX}, 000--000}
   \setcounter{page}{1}

   \author{Yu-Yang Li\inst{1}, Kai Li\inst{1}, Yuan Liu\inst{2}
   }

   \institute{ Shandong Provincial Key Laboratory of Optical Astronomy and Solar-Terrestrial Environment, Institute of Space Sciences, Shandong University, Weihai, 264209, China; {\it kaili@sdu.edu.cn}\\
\and
Qilu Institute of Technology, Jinan 250200, China\\
\vs \no
}

\abstract{V0842 Cep is a W UMa-type binary star that has been neglected since its discovery. We analysed the VR$_c$I$_c$ light curves, obtained by the 1 m telescope at the Weihai Observatory of Shandong University, using the Wilson-Devinney code. V0842 Cep was found to be a shallow contact binary system (f=8.7$\%$) with a mass ratio of 2.281. Because its orbital inclination is greater than 80$^\circ$, the photometric results are reliable. A period study is included which reveals a continually decreasing orbital period ($\frac{\mathrm{d}p}{\mathrm{d}t}$=1.50($\pm$0.42)$\times$10$^{-7}$dyr$^{-1}$). This trend could be attributed to the angular momentum loss via stellar wind.
\keywords{stars: binaries: close --- stars: binaries: eclipsing --- stars: individual (V0842 Cep)
}
}

   \authorrunning{Li et al. }            
   \titlerunning{The first photometric analysis and period investigation of V0842 Cep}  
   \maketitle

%
\section{Introduction}           
\label{sect:intro}
W UMa-type binary systems (contact binaries) usually contain two main sequence stars whose spectral types vary from F to K. There is a common convective envelope that encases both stars. The orbital periods of these systems are generally less than a day. In the past fifty years, many researchers have investigated these types of binaries (e.g., \citealt{Hilditch89,Stepien06,Zhang11,Qian13a,Qian14,Qian17}); however, there are still a few questions regarding their structures and evolution, such as short-period cutoff (e.g., \citealt{Rucinski92,Jiang12,Li19,Li20}); and lower mass-ratio limit(e.g., \citealt{Webbink76,Yang15,Kjurkchieva18,Kjurkchieva20}); Therefore, W UMa-type binaries should be further investigated.

V0842 Cep is a typical W UMa-type eclipsing binary, which exhibits continuous light variation, and the depths of two light minima are similar. The orbital period of V0842 Cep (GSC 04586-01412,V=14.7mag) is 0.288957 days,making it a very short period system (VSP). The object was discovered by \citet{Kuzmin07} and confirmed by ASAS-SN\nolinebreak\footnotemark[1] \footnotetext[1]{http://www.astronomy.ohio-state.edu/asassn/index.shtml.} \citep{Shappee14,Jayasinghe19}. \cite{Jayasinghe19} found that this object is an EW-type binary with V magnitude of 14.34 mag, amplitude of 0.67 mag, and period of 0.288954 days. Using the observations of VR$_c$I$_c$ charge-coupled device (CCD) light curves, we include here a complete synthetic light curve analysis and period study.

\section{Observation}
\label{sect:Obs}

V0842 Cep was observed using the 1 m telescope at Weihai Observatory, Shandong University \citep{Hu14} on October 24, 2018. To this end, the PIXIS 2048B CCD camera was also employed; the camera has a format of 2048 $\times$ 2048 pixels, revealing a 12$^{'}$$\times$12$^{'}$ field of view. The Johnson--Cousin--Bessel VR$_c$I$_c$ data were derived using the CCD photometric system. The typical exposure time is 100 s for the V band, 45 s for the R band, and 25 s for the I band.

The data were reduced using C-Munipack\nolinebreak\footnotemark[2] \footnotetext[2]{C-Munipack is a CCD photometry data processing software, developed based on the Munipack. The program supports graphical user interfaces and command line models.} including bias and flat correction. We then used an aperture photometry method to measure instrumental magnitudes. Figure 1 shows one of the processed images, where the variable star (i.e., V0842 Cep), comparison star, and check star are denoted as "V", "C", and "CH", respectively. The photometric results of the three bands as well as the difference in the magnitudes of the comparison and check stars are shown in Figure 2. Two light minima were determined as 2458415.99495($\pm$0.00056) and 2458416.13833($\pm$0.00054).
\begin{figure}
\centering
\includegraphics[width=8cm, angle=0]{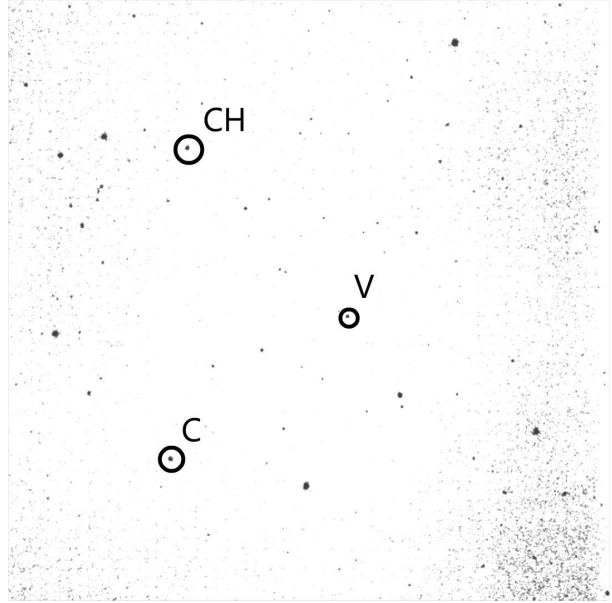}
\caption{CCD image around V0842 Cep has been presented. The "V" denotes the variable star; the "C", comparison star; and "CH", check star.}
\label{Figure1}
\end{figure}

\begin{figure}
\centering
\includegraphics[width=8cm, angle=0]{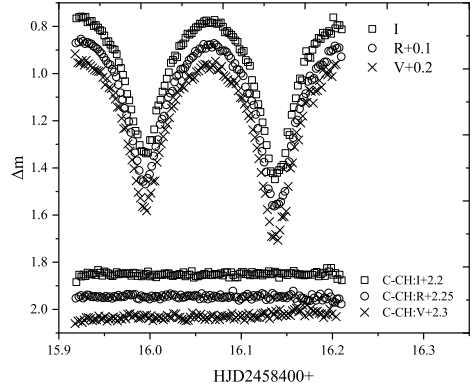}
\caption{Light curves of V0842 Cep at three bands. The figure also presents the difference in the magnitude between the comparison and the check stars.}
\label{Figure2}
\end{figure}

\section{Photometric solutions}

We analysed the VR$_c$I$_c$ light curves via the W-D code \citep{Wilson71,Wilson90,Wilson94}. First, we determined stellar effective temperature 4920 K from Gaia DR2 database\nolinebreak\footnotemark[3] \footnotetext[3]{https://gea.esac.esa.int/archive/.} \citep{Gaia16,Gaia18}, and noticed that its typical accuracy is 324K \citep{Andrae18}. Preliminary solutions demonstrate that the luminosity of the secondary star is greater than that of the primary star; therefore, we set the stellar effective temperature as the temperature of the secondary star. V0842 Cep is a late-type binary star; thus, a few parameters can be fixed to match the corresponding convective envelope, such as the gravity-darkening coefficients (g$_1$ =g$_2$ =0.32) \citep{Lucy67} and bolometric albedo (A$_1$ = A$_2$ = 0.5) \citep{Rucinski69}. Bolometric and bandpass limb-darkening coefficients can be obtained from \citet{Van93}.

\begin{figure}
\centering
\includegraphics[width=8cm, angle=0]{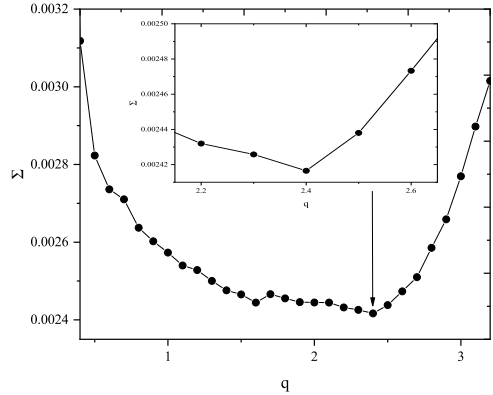}
\caption{Relationship between q and $\Sigma$, and the inserted figure presents the values around the lowest point.}
\label{Figure3}
\end{figure}
Evidently, the three-bandpass light curves are EW-type, indicating that V0842 Cep is a contact binary system. Therefore, we selected mode 3 during our modelling. There are a few adjustable parameters, such as the monochromatic luminosity of the primary component (L$_{1V}$, L$_{1R}$ and L$_{1I}$), orbital inclination (i), effective temperature of the primary component (T$_{1}$), and potentials of two components ($\Omega$$_{1}$ =$\Omega$$_{2}$).

We determined the mass ratio of V0842 Cep via a q-search method. We plotted the relation between mass ratio q and the weighted sum of the squared residuals $\Sigma$W$_i$(O-C)$_i$ in Figure 3 to obtain the most reliable mass ratio. When q is 2.4, the $\Sigma$ can be employed to derive the minimum value. Therefore, we set q = 2.4 as an initial value and ensured that it could be varied. Then, the results were derived. Photometric solutions are summarised in Table 1 and the corresponding theoretical light curves are illustrated in Figure 4.
\begin{table}
\bc
\caption[]{Photometric solutions of the V0842 Cep.\label{tab1}}\
\setlength{\tabcolsep}{1pt}
\small
 \begin{tabular}{ccc}
  \hline\noalign{\smallskip}
Parameters &                   Value&                    Errors      \\
\hline
     $g_1=g_2$ &               0.32 &                   Assumed     \\
     $A_1=A_2$ &               0.5  &                   Assumed     \\
     $T_1(K)$  &               5197 &                   12     \\
     $T_2(K)$  &               4920 &                   Assumed     \\
     $q(M_2/M_1)$ &            2.281&                 0.034     \\
     $\Omega_{in}$ &           5.6450&                    Assumed        \\
     $\Omega_{out}$ &          5.0401&                     Assumed        \\
     $\Omega_1$=$\Omega_2$ &   5.5926&                 0.0428     \\
     $i$ &                     80.0&                0.2     \\
     $(L_1/{L_1+L_2})_V$&         0.3944&                 0.0024     \\
     $(L_1/{L_1+L_2})_R{_c}$&         0.3796&                 0.0022     \\
     $(L_1/{L_1+L_2})_I{_c}$&         0.3704&                 0.0021     \\
     $r_1(pole)$ &             0.2938&                 0.0010     \\
     $r_1(side)$ &             0.3070&                 0.0011     \\
     $r_1(back)$ &             0.3425&                 0.0014     \\
     $r_2(pole)$ &             0.4295&                 0.0044     \\
     $r_2(side)$ &             0.4584&                 0.0059     \\
     $r_2(back)$ &             0.4873&                 0.0082     \\
     $f$&                      8.7$\%$&                  7.1$\%$     \\
  \noalign{\smallskip}\hline
\end{tabular}
\ec
\end{table}
\begin{figure}
\centering
\includegraphics[width=8cm, angle=0]{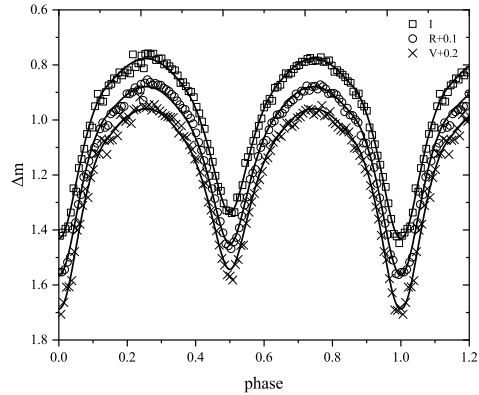}
\caption{Fitting result of the theoretical and observational light curves.}
\label{Figure4}
\end{figure}

\section{Orbital period investigations}
\label{sect:Obs}

In addition to the results calculated using our own data and AAVSO\nolinebreak\footnotemark[4] \footnotetext[4]{https://www.aavso.org/}, the light minimum times can also be obtained from the O-C Gateway\nolinebreak\footnotemark[5] \footnotetext[5]{http://var.astro.cz/ocgate/.} and relevant references.

There are a total of 8 times of light minimum, and the corresponding O-C values were calculated by a linear ephemeris:
\begin{equation}
Min.I=2456908.3776+0.288957E.
\end{equation}
The O-C curve can be fitted to represent the dotted line in Figure 5; however, it cannot completely describe the variation trend. Therefore, we used the least-squares method and found a good fit by adding a quadratic term:
\begin{eqnarray}\label{equ:2}
Min.I&=&2456908.37367(\pm0.00295)+0.28895094(\pm0.000000432171)\times E\\\nonumber
&+&-5.95(\pm1.65)\times10^{-11}E^2.
\end{eqnarray}
Based on the quadratic term, we could calculate the period decrease rate: $\frac{\mathrm{d}p}{\mathrm{d}t}$=1.50($\pm$0.42$)\times$10$^{-7}$dyr$^{-1}$.
The fitting curve is plotted and illustrated as the solid line in Figure 5 further, the corresponding residuals are presented in the bottom of Figure 5. The details of each data point are listed in Table 2.
\begin{figure}
\centering
\includegraphics[width=8cm, angle=0]{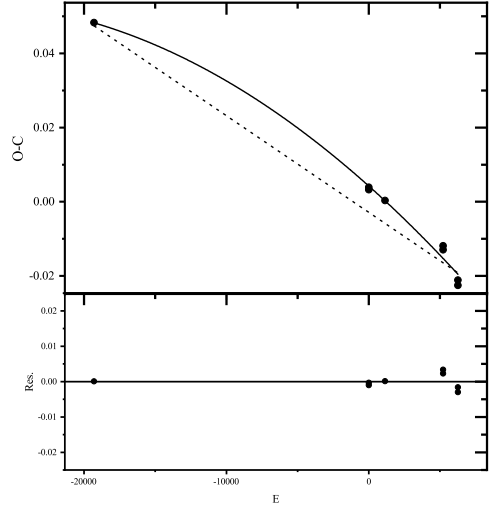}
\caption{O-C diagram of V0842 Cep.}
\label{Figure5}
\end{figure}

\section{Discussion}
\label{sect:discussion}

The VR$_c$I$_c$ light curves were analysed via the W-D program. This is the first to determine the photometric solutions of V0842 Cep. The solutions indicate that V0842 Cep is a W-type shallow contact binary with a mass ratio of 2.281 and contact degree of 8.7$\%$. And its orbital inclination is more than 80 degrees, which means the photometric solution results are quite reliable \citep{Pribulla03,Terrell05}. Since there are no spectroscopic observations of this star, absolute parameters cannot be uniquely determined. If the more massive component is a normal, main sequence star, its estimated mass is 0.76M$_{\odot}$ according to the effective temperature 4920 K \citep{Cox00}. The mass of the less massive component can be calculated as follows: q=M$_2$/M$_1$. The result obtained is 0.33M$_{\odot}$. Other absolute parameters are a=1.89R$_{\odot}$, R$_1$=0.59R$_{\odot}$, R$_2$=0.87R$_{\odot}$, L$_1$=0.23L$_{\odot}$, and L$_2$=0.40L$_{\odot}$.

According to the colour index obtained from APASS-DR9\nolinebreak\footnotemark[6] \footnotetext[6]{The AAVSO Photometric All Sky Survey (APASS) project is designed to bridge the gap between the shallow Tycho2 two-bandpass photometric catalogue that is complete to V=11 and the deeper \citep{Henden16}.} and interstellar reddening coefficient measured by \citet{Schlafly11}, the (B-V)$_0$ can be determined as 0.6238. Using the following equation proposed by \citet{Rucinski97}, the absolute magnitude was calculated to be M$_V$ = 4.40 mag:
\begin{equation}
M_V=-4.44logP+3.02(B-V)_0+0.12,
\end{equation}
According to \citet{Samus17}, the V band magnitude under maximum light can be determined to be m$_V$ = 14 mag; further, after considering interstellar extinction, the revised value can be calculated as 12.80 mag \citep{Schlafly11}. The distance module can then be determined as m$_V$ - M$_V$ =8.40 mag, and the corresponding distance can be estimated to be 478.63 pc, which is similar to the distance derived by Gaia DR2, 485.12 pc.

To estimate the evolutionary status of V0842 Cep, both the components of V0842 Cep are denoted on the mass-luminosity diagram in Figure 6. We constructed Zero Age Main Sequence (ZAMS) and Terminal Age Main Sequence (TAMS) lines using the binary star evolution code provided by \citet{Hurley02}. Simultaneously, the parameters of 42 W-type binaries, obtained by \citet{Yakut05}, are indicated for comparison. The primary star of the object is located above the TAMS, implying that the primary star has evolved out of the main sequence and is over-luminous and over-sized. In contrast, the secondary star of the object lies between the ZAMS and the TAMS, indicating that it is an evolved main sequence star. V0842 Cep exhibits an evolutionary status similar to those of other W-type binaries.

\begin{figure}
\centering
\includegraphics[width=8cm, angle=0]{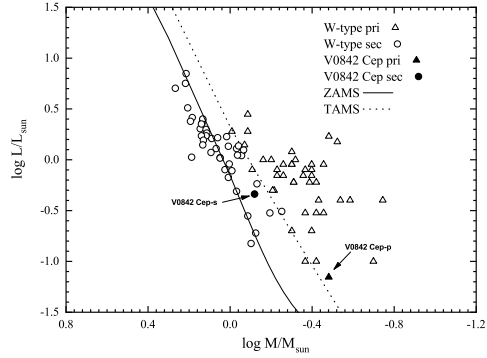}
\caption{Log M-Log L diagram: Zero Age Main Sequence (ZAMS) and Terminal Age Main Sequence (TAMS) lines are constructed based on the binary star evolution code provided by Hurley et al. (2002).  Simultaneously, the observations of 42 W-subtype binaries, obtained by Yakut \&Eggleton (2005), are indicated for comparison. Both components of V0842 Cep are denoted using bold arrows.}
\label{Figure6}
\end{figure}
\begin{table}
\bc
\begin{minipage}[]{100mm}
\caption[]{Times of Light Minimum.\label{table2}}\end{minipage}
\setlength{\tabcolsep}{2.5pt}
\small
 \begin{tabular}{ccccccccccccc}
  \hline\noalign{\smallskip}
 JD (Hel.) & Error & Method &  Type & E & (O-C)$_1$ & (O-C)$_2$ & Residuals & Reference     \\ \hline
2451330.6850 &        & ccd & P & -19303 & -0.00255 & -0.09122 & 0.0001 & O-C gateway  \\
2456908.3776 & 0.0003 & ccd & P & 0 & 0.00393 & 0.0036 & -0.0003 & OEJV 0179  \\
2456908.5214 & 0.0004 & ccd & S & 0.5 & 0.00325 & 0.00292 & -0.0010 & OEJV 0179  \\
2457238.5074 & 0.0003 & ccd & S & 1142.5 & 0.00337 & 0.00688 & 0.0001 & OEJV 0179  \\
2458715.7786 & 0.0008 & ccd & P & 6255 & -0.00404 & 0.01415 & -0.0016 & This paper (AAVSO)  \\
2458716.7885 & 0.0005 & ccd & S & 6258.5 & -0.00605 & -0.01274 & -0.0030 & This paper (AAVSO)  \\
2458415.9950 & 0.0006 & ccd & S & 5217.5 & 0.00188 & 0.01781 & 0.0034 & This paper (1m)  \\
2458416.1383 & 0.0005 & ccd & P & 5218 & 0.00078 & 0.1672 & 0.0023 & This paper (1m)  \\
  \noalign{\smallskip}\hline
\end{tabular}
\ec
\end{table}

According to the 8 minimum light times, we did the period analysis for the first time. The downward parabolic trend of O-C means a long-term decrease in $\frac{\mathrm{d}p}{\mathrm{d}t}$=1.50($\pm$0.42)$\times$10$^{-7}$d yr$^{-1}$. Assuming this trend is caused by the conservative mass transfer between two components, corresponding parameters should satisfy the following equation:
\begin{equation}
\frac{\dot{P}}{P}=-3\dot{M_1}(\frac{1}{M_1}-\frac{1}{M_2}),
\end{equation}
We can then obtain the following: $\frac{\mathrm{d}M_1}{\mathrm{d}t}$=+1.01($\pm$0.28)$\times$10$^{-7}$M$_{\odot}$ yr$^{-1}$. The positive sign implying that the more massive component is losing mass, and the corresponding timescale should be $\tau$$\sim$$\frac{M_2}{\dot{M_2}}$$\sim$7.52$\times$10$^{6}$yr, which is approximately 7 times shorter than the thermal timescale of the more massive component ($\frac{GM^2}{RL}$$\sim$5.20$\times$10$^{7}$)yr. Therefore, the long-term period decreases may be attributed to the angular momentum loss via magnetic stellar wind. By the decreasing orbital period, we can estimate that the shallow contact binary system will evolve into a deep contact binary system in the future.

V0842 Cep is a K-type binary. Thus, a few K-type contact binaries have been listed and compared in Table 3. We statistically determined that K-type binaries have a few common features. First, temperature of the more massive component is lower than that of the primary component for all these systems, indicating that most of the K-type contact binaries are W-subtype systems. Second, most systems are shallow contact binaries(f$\le$25$\%$), corresponding to the character of W-subtype contact binaries. Third, the periods of these systems are fairly short (less than 0.3 days), which can be satisfied with the period--colour relation \citep{Rucinski98}.

V0842 Cep is worthy of further investigation. More spectroscopic and photometric observations are required for determining the precise mass ratio and orbital period variation of binary stars.
\begin{table}
\bc
\begin{minipage}[]{100mm}
\caption[]{Parameters for K-type binaries.\label{table3}}\end{minipage}
\setlength{\tabcolsep}{2.5pt}
\small
 \begin{tabular}{ccccccccccccc}
  \hline\noalign{\smallskip}
Binary & period(d) & q & f & T$_1$(K) & T$_2$(K)& $\frac{\mathrm{d}p}{\mathrm{d}t}$ & Reference  \\ \hline
           NSVS 4484038 & 0.2186 & 2.74 & 10$\%$ & 4839 & 4750 & - & \citet{Zhang14}  \\
           CC Com & 0.2207 & 1.90 & 17$\%$ & 4300 & 4200 & -1.3$\times$10$^{-8}$ & \citet{Kose11}  \\
           07g-3-00820 & 0.2270 & 1.30 & 5$\%$ & 5100 & 4967 & - & \citet{Gao17}  \\
           V1104 Her & 0.2279 & 1.61 & 17.1$\%$ & 4050 & 3902 & -2.9$\times$10$^{-8}$ & \citet{Liu15}  \\
           1SWASP J161335.80-284722.2 & 0.2298 & 1.06 & 18.7$\%$ & 4317 & 4126 & -4.3$\times$10$^{-7}$ & \citet{Fang19}  \\
           V523 Cas & 0.2337 & 1.76 & 21.6$\%$ & 5082 & 4763 & - & \citet{Jeong10}  \\
           YZ Phe & 0.2347 & 2.64 & 9.7$\%$ & 4908 & 4658 & -2.6$\times$10$^{-8}$ & \citet{Samec95}  \\
           FY Boo & 0.2411 & 2.55 & 11$\%$ & 4750 & 4555 & - & \citet{Samec11}  \\
           NSVS 2706134 & 0.2448 & 2.60 & 15.3$\%$ & 4660 & 4204 & - & \citet{Martignoni16}  \\
           1SWASP J064501.21+342154.9 & 0.2486 & 2.11 & 15.3$\%$ & 4450 & 4367 & - & \citet{Liu14b}  \\
           BI Vul & 0.2518 & 1.04 & 4.4$\%$ & 4600 & 4474 & -9.5$\times$10$^{-8}$ & \citet{Qian13b}  \\
           FS Cra & 0.2636 & 1.32 & 14.6$\%$ & 4700 & 4567 & - & \citet{Bradstreet85}  \\
           IL Cnc & 0.2677 & 1.76 & 8.9$\%$ & 5000 & 4731 & 7.0$\times$10$^{-9}$ & \citet{Liu20}  \\
           RV CVn & 0.2696 & 1.74 & 9.8$\%$ & 4750 & 4607 & - & \citet{Liu14a}  \\
           FG Sct & 0.2706 & 1.35 & 21.4$\%$ & 4536 & 4373 & 6.4$\times$10$^{-8}$ & \citet{Yue19}  \\
           V0842 Cep & 0.2890 & 2.28 & 8.7$\%$ & 5197 & 4920 & -1.5$\times$10$^{-7}$ & This paper  \\
           V1799 Ori & 0.2903 & 1.34 & 3.5$\%$ & 5000 & 4781 & 1.8$\times$10$^{-8}$ & \citet{Liu14c}  \\
  \noalign{\smallskip}\hline
\end{tabular}
\ec
\end{table}
\normalem
\begin{acknowledgements}

This work is supported by National Natural Science Foundation of China (NSFC) (No. 11703016), and by the Joint Research Fund in Astronomy (No. U1931103) under cooperative agreement between NSFC and Chinese Academy of Sciences (CAS), and by the Natural Science Foundation of Shandong Province (Nos. ZR2014AQ019), and by Young Scholars Program of Shandong University, Weihai (Nos. 20820171006), and by Chinese Academy of Sciences Interdisciplinary Innovation Team, and by the Open Research Program of Key Laboratory for the Structure and Evolution of Celestial Objects (No. OP201704).

We would like to acknowledge the AAVSO International Database (https://www.aavso.org/), from which variable star observations were obtained. Observers worldwide have contributed to this database. Further, this article uses data from the All-Sky Automated Survey for SuperNovae \citep{Shappee14,Jayasinghe19}, which consists of two telescopes on a common mount and is hosted by the Las Cumbres Observatory Global Telescope Network in the Faulkes Telescope North enclosure on Mount Haleakala, Hawaii. Finally, this work has utilised data listed in the European Space Agency (ESA) mission Gaia (https://www.cosmos.esa.int/gaia), processed by the Gaia Data Processing and Analysis Consortium (DPAC,https://www.cosmos.esa.int/web/gaia/dpac/consortium). Funding for the DPAC was provided by national institutions, in particular the institutions participating in the Gaia Multilateral Agreement.

This work is partly supported by the Supercomputing Center of Shandong University, Weihai.
\end{acknowledgements}

\label{lastpage}
\end{document}